\documentclass[twocolumn,secnumarabic,amssymb, nobibnotes, aps, prd]{revtex4}

\usepackage{graphicx}
\usepackage{dcolumn}
\usepackage{bm}
\usepackage{ulem}

\begin{document}

\title{Collisional Properties of p-Wave Feshbach Molecules}

\author{Yasuhisa Inada$^{1, 2}$, Munekazu Horikoshi$^{1}$, Shuta Nakajima$^{1, 3}$, Makoto Kuwata-Gonokami$^{1, 2}$, Masahito Ueda$^{1, 3}$}
\author{Takashi Mukaiyama$^{1}$\footnote{Email address: muka@sogo.t.u-tokyo.ac.jp}}
\affiliation{$^{1}$\mbox{ERATO Macroscopic Quantum Control Project, JST, Yayoi, Bunkyo-Ku, Tokyo 113-8656, Japan}\\
$^{2}$\mbox{Department of Applied Physics, University of Tokyo, Hongo, Bunkyo-Ku, Tokyo 113-8656, Japan}\\
$^{3}$\mbox{Department of Physics, University of Tokyo, Hongo, Bunkyo-ku, Tokyo 113-0033, Japan}}

\date{\today}
\begin{abstract}
We detected the formation of $p$-wave Feshbach molecules for all three combinations of the two lowest atomic spin states of $^6$Li. By creating a pure molecular sample in an optical trap, we measured the inelastic collision rates of $p$-wave molecules. The elastic collision rate was measured from the thermalization rate of a breathing mode which was excited spontaneously upon molecular formation. 
\end{abstract}
\maketitle

Ultracold atomic gases with tunable interaction via Feshbach resonance have opened up possibilities for studying a variety of novel phenomena in quantum degenerate phases. 
In particular, $s$-wave Feshbach resonances have been widely used to study the crossover between  Bose-Einstein condensation (BEC) and Bardeen-Cooper-Schrieffer (BCS) superfluidity\cite{Regal1,Zwierlein,Bourdel,Chin,Kinast,Partridge}.
One of the greatest challenges in the field of trapped Fermi gases is the realization of superfluidity with pairs of atoms in a non-zero relative orbital angular momentum state\cite{Levinsen,Bohn,You}. In such systems, a rich variety of phases are expected to merge due to complex superfluid order parameters\cite{SadeMelo,Ohashi,Ho,Gurarie,Cheng}. Non-$s$-wave superfluidity has so far been realized in superfluid $^{3}$He ($p$-wave)\cite{He}, strontium ruthenate superconductors ($p$-wave)\cite{Maeno}, and high-$T_{\rm{c}}$ copper-oxide superconductors ($d$-wave).

Until now, $^{40}$K and $^6$Li have been utilized to study $p$-wave Feshbach resonances\cite{Zhang,Fuchs,Chevy,Regal2,Gaebler,Jin,Gunter,Schunck}. 
The positions of $p$-wave Feshbach resonances of $^6$Li have been accurately determined by Zhang {\it et al.}\cite{Zhang} and Schunck {\it et al.}\cite{Schunck}.
Zhang {\it et al.} also detected the formation of $p$-wave Feshbach molecules of $^6$Li in one of the three combinations of the two lowest atomic hyperfine spin states, $|F=1/2,m_F=1/2\rangle$ ($\equiv |1\rangle$) and $|F=1/2,m_F=-1/2\rangle$ ($\equiv |2\rangle$), but molecules were not detected in the $|1\rangle - |1\rangle$ state, which is the most promising candidate for a $p$-wave condensate because the dipolar loss is energetically suppressed.
Recently, Fuchs {\it et al.} measured the binding energies and relative molecular magnetic moments of $|1\rangle - |1\rangle$, $|1\rangle - |2\rangle$, and $|2\rangle - |2\rangle$ molecules of $^6$Li\cite{Fuchs}. 
Gaebler {\it et al.} successfully created $p$-wave Feshbach molecules of $^{40}$K and observed the anisotropic energy release at the dissoication process\cite{Gaebler,Jin}. The lifetime of $^{40}$K molecules in both $m_l =\pm 1$ and $m_l =0$ states were found to be less than 10 ms, even on the bound side of the resonance due to dipolar relaxation into lower atomic spin states\cite{Jin}.

\begin{figure}[b]
\includegraphics[scale = 0.55]{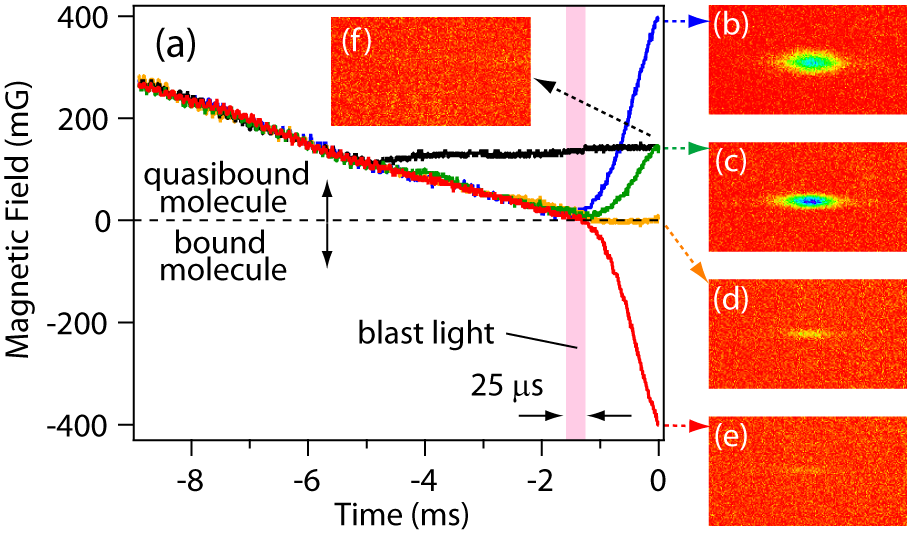}\\
\caption{(a) Time sequence of magnetic field to create molecules. Magnetic field is ramped adiabatically from above toward the Feshbach resonance (dashed line). (b, c) Absorption images of dissociated molecules by the upward magnetic field ramp. (d, e) Atoms in a molecular gas detected at (d) or below (e) the resonance. (f) No atoms and molecules are detected when the magnetic field is off resonance.\label{sequence}}
\end{figure}

While $s$-wave Feshbach molecules are highly stable against vibrational quenching near a Feshbach resonance,
it is not clear how much $p$-wave molecules would suffer from inelastic collisions due to an intrinsically small interatomic separation caused by the centrifugal barrier. Therefore, studying collisional properties of $p$-wave Feshbach molecules is an important step toward realization of $p$-wave molecular condensates. In this Letter, we present elastic and inelastic collisional properties of $^6$Li$_2$ $p$-wave Feshbach molecules.
To suppress the loss of molecules, we irradiated a resonant light pulse and blew away the residual atoms from the trap immediately after the molecules were created\cite{Xu,Mark,Syassen}. The resultant pure molecular sample in the optical trap enables us to measure inelastic decay rates of molecules precisely. We measured the atom-dimer inelastic collision coefficients by keeping both atoms and molecules in the trap, and dimer-dimer inelastic collision coefficients by using a pure molecular sample. We also measured the dimer-dimer elastic collision rate from the thermalization rate of breathing-mode oscillations of the cloud, which were induced spontaneously by the mismatch between the initial size of the atomic cloud and the equilibrium size of the molecular cloud\cite{Mukaiyama}.

\begin{figure}[t]
\includegraphics[scale = 0.7]{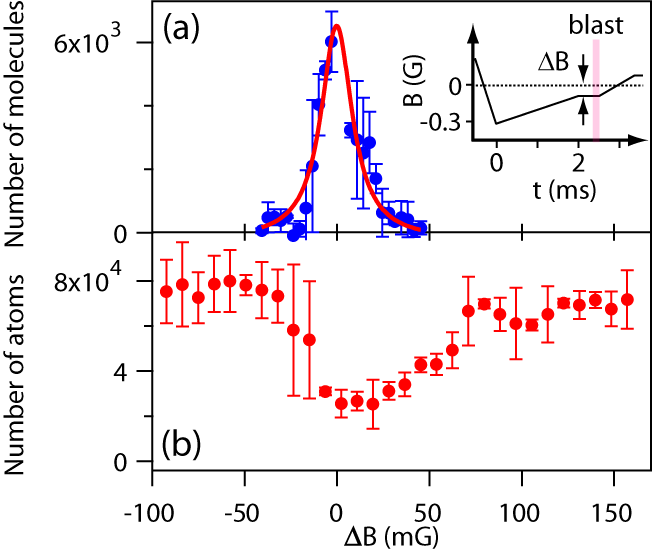}\\
\caption{(a) Number of $|1\rangle - |1\rangle$ molecules created as a function of magnetic field detuning $\Delta B$. The solid curve shows a Lorentzian fit to the data. Inset shows the time sequence of magnetic field for data taking. (b) Number of atoms remaining after a hold time of 25~ms as a function of $\Delta B$.
\label{creation}}
\end{figure}

Our experiments were performed using a quantum degenerate gas of $^6$Li in the $|1\rangle$ and $|2\rangle$ spin states. In the present experiment, we evaporated the atoms in a crossed beam optical dipole trap at 300~G, where the elastic collision cross section between $|1\rangle$ and $|2\rangle$ takes on a local maximum. 
After evaporation for 4~s, the mixture of $5 \times 10^5$ atoms in $|1\rangle$ and $|2\rangle$ with equal population and the temperature of $T/T_{\rm{F}}\sim$~0.25 was prepared. 
The optical trap beams were orthogonal to each other on the horizontal plane: One, oriented along the $x$ axis, has beam waist of 26~$\mu$m and the other, oriented along the $y$ axis, has an ellipsoidal shape with a waist of 360~$\mu$m in the $x$ direction and 170~$\mu$m in the $z$ (vertical) direction. The trap frequencies ($\omega_{x}$, $\omega_{y}$, $\omega_{z}$) at the final trap depth are $2 \pi$  $\times$ (39, 760, 780) Hz. By stabilizing the currents in the Feshbach coils, we suppressed magnetic-field fluctuations due to current noise down to 10~mG.

To create $|1\rangle - |1\rangle$ molecules, we first ramp the magnetic field up to the Feshbach resonance of $|2\rangle - |2\rangle$ at 215~G to eliminate atoms in the $|2\rangle$ state. 
After preparing the atoms in the pure $|1\rangle$ state, we create molecules by ramping the magnetic field to a value near the Feshbach resonance of the $|1\rangle - |1\rangle$ state at 159~G. 
Due to the smallness of separation of Feshbach resonances for $m_l =\pm 1$ and $m_l =0$ molecules\cite{Chevy}, we were unable to select a specific $m_l$ state.
The time sequence used to create $|1\rangle - |1\rangle$ molecules is shown in Fig. \ref{sequence}(a). 
When the magnetic field reached the resonance, the sample was irradiated with a 25-$\mu$s pulse of resonant light (``blast'') to remove unpaired atoms from the trap. 
Molecules are relatively immune to the blast light due to the smallness of the Franck-Condon factor for the optical transition.
After blasting the atoms, the magnetic field was ramped up above the resonance (Fig. \ref{sequence}(b,c)), kept at the resonance (Fig. \ref{sequence}(d)), or ramped down below the resonance (Fig. \ref{sequence}(e)) before taking absorption images. The blue, green, orange, and red curves in Fig. \ref{sequence}(a) correspond to the magnetic-field sequence for the images in Fig. \ref{sequence}(b-e), respectively. Ramping the magnetic field to a value above the resonance as shown by the blue and green curves in Fig. \ref{sequence}(a), allows us to clearly image the dissociated atoms(Fig. \ref{sequence}(b,c)). As shown by the orange and red curves in Fig. \ref{sequence}(a), the atoms in bound molecules are detected, albeit weakly, because of a finite Franck-Condon factor for the optical transition.
When the magnetic field is kept away from the resonance by $\Delta B = + 100$ mG, no atoms were detected (Fig. \ref{sequence}(f)), indicating the blast light completely removed the unpaired atoms.
Therefore, the atoms imaged in Fig. \ref{sequence}(d,e) are not unpaired atoms but atoms that form molecules.
As seen in Fig. \ref{sequence}(b-d), the higher the magnetic field was applied for dissociation, the more widely the cloud expanded during the same time of flight of 500~$\mu$s. This indicates that faster ramping of the magnetic field leads to higher dissociation energy.
We produced $|1\rangle - |2\rangle$  and $|2\rangle - |2\rangle$ molecules in a similar manner by using Feshbach resonances at 185 and 215~G, respectively.

\begin{figure}[b]
\includegraphics[scale = 0.55]{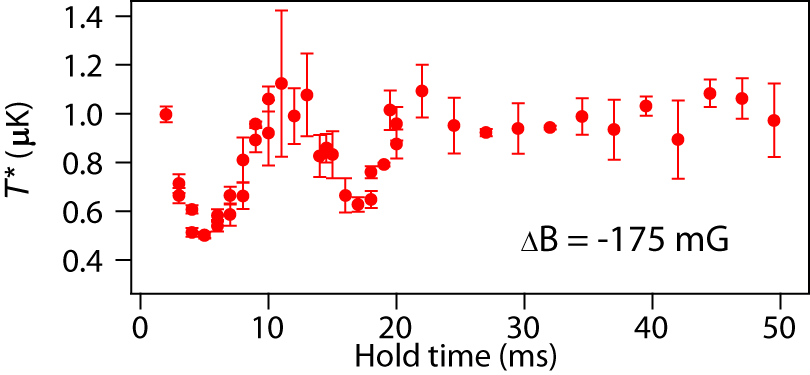}
\caption{Time evolution of $T^*$. Thermalization time is estimated to be 20~ms.\label{oscillation}}
\end{figure}

To investigate the range of magnetic field over which molecules can be formed efficiently, we ramped the magnetic field from below up to the resonance. This avoided the adiabatic formation of molecules. As shown in the inset of Fig. \ref{creation}(a), we first ramped the field across the resonance at a fast rate in order to minimize the atomic loss. We then swept the magnetic field in 1~ms to a value at which the number of molecules formed was measured. After molecules were formed during a hold time of 500~$\mu$s, the blast light was turned on to remove unpaired atoms. Figure \ref{creation}(a) shows the number of molecules formed vs magnetic-field detuning $\Delta B$, where $\Delta B$ is defined as positive when the magnetic field is higher than the resonance and the molecular state becomes quasibound. Throughout the paper, $\Delta B=0$ is defined as the magnetic field at which the maximum number of molecules is created. At the peak, we can convert 15 \% of atoms to molecules, which is consistent with previous experiments\cite{Gaebler,Zhang}.
 Figure \ref{creation}(b) shows the atomic loss measured after the 25-ms hold time for each magnetic field. An asymmetry in the loss feature is clearly seen, as observed by other groups\cite{Schunck,Fuchs}. Note that the bottom of the loss curve does not coincide with the peak of the molecular creation curve. The physical origin of this differecne is unclear and requires further study.

In the process of creating molecules, breathing oscillations were excited spontaneously. This is because the molecules were created over the entire cloud of the Fermi-degenerate atoms, which is larger than the equilibrium cloud size of molecules\cite{Mukaiyama}. Figure \ref{oscillation} shows the time evolution of $T^*(t)$, defined by $T^*(t)=m \omega^2 \sigma_{\rm{rms}}^2(t)/k_{\rm{B}}$, where $m$ is the mass of molecules, $\omega$ is $2 \pi$ times the trap frequency along the longest axis ($x$ axis) of the trap, and $\sigma_{\rm{rms}}$ is the root-mean-square size of the molecular cloud.
The size of the molecular cloud is largest when the molecules are created, and starts to shrink due to the optical trapping potential. The frequency of initial oscillations is evaluated to be 87(10)~Hz, which agrees with twice the trap frequency of 39~Hz in the $x$ direction. As seen in Fig. \ref{oscillation}, $T^*(t)$ shows abrupt thermalizaion at $t=20$~ms, by which we determined the thermalization time to be 20~ms. It is reported that it takes about 2.7 collisions for the cloud to thermalize\cite{Monroe,Wu}. Based on this, the dimer-dimer elastic collision rate is estimated to be $(20 \rm{ms}/2.7)^{-1} \sim 140$~Hz.

\begin{figure}[tb]
\includegraphics[scale = 0.65]{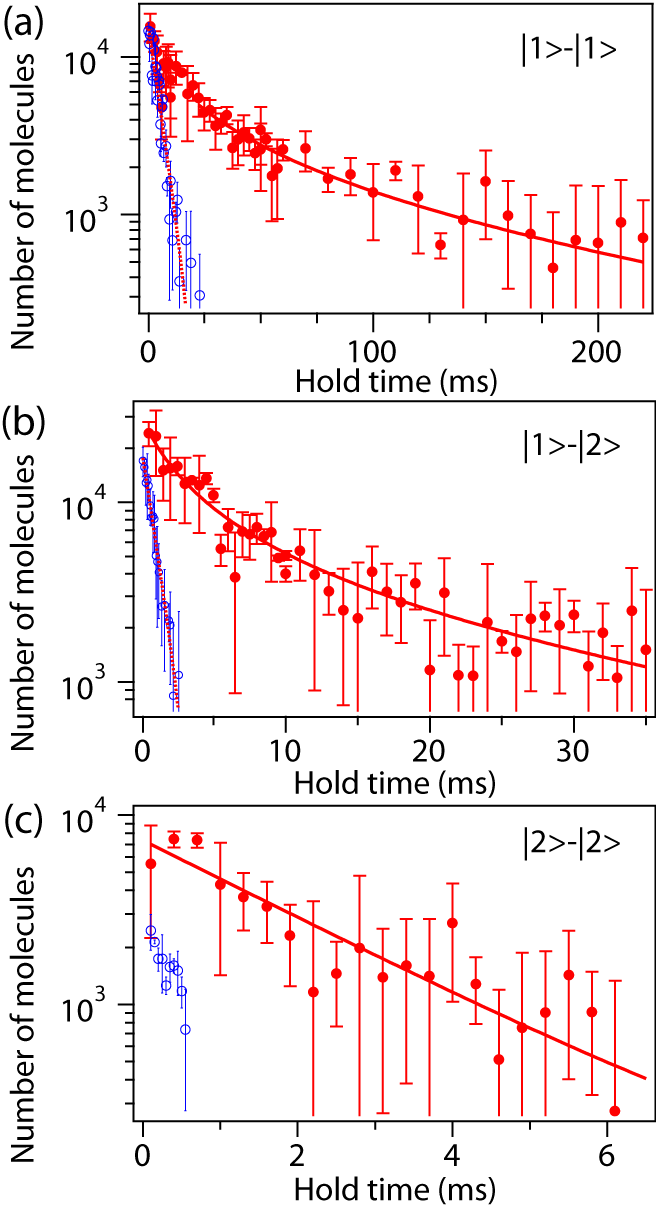}
\caption{Number of molecules vs hold time for $|1\rangle - |1\rangle$ (a), $|1\rangle - |2\rangle$ (b), and $|2\rangle - |2\rangle$ molecules (c). Filled circles show the decay data for molecules alone, and open circles show the decay data with both atoms and molecules. The solid and dotted curves are for pure molecules and for atom-molecule mixtures.\label{lifetime}}
\end{figure}

We measured the lifetimes of molecules in all three combinations of $|1\rangle$ and $|2\rangle$ states (Fig. \ref{lifetime}). Figures \ref{lifetime}(a)-\ref{lifetime}(c) show respectively the numbers of $|1\rangle - |1\rangle$, $|1\rangle - |2\rangle$, and $|2\rangle - |2\rangle$ molecules remaining after the varying hold time. 
In \ref{lifetime}(a), the number of molecule is greater than that in Fig. \ref{creation}(a) due to different experimental conditions.
Filled circles show the number of molecules vs hold time under the condition that only molecules are held in the trap, and open circles show the same quantity in the presence of unpaired atoms. In the latter case, the blast light was turned on after the hold time to ensure that the signal arises exclusively from molecules. The molecular lifetime is greatly suppressed in the presence of unpaired atoms. Magnetic-field detuning during the hold time is $\Delta B = -175 \rm{mG}$ for all data points.

The loss of molecules is analyzed using the following equation\cite{Syassen}
\begin{equation}
\frac{\dot n_{\rm{d}}}{n_{\rm{d}}}= -\alpha - K_{\rm{ad}} n_{\rm{a}} - K_{\rm{dd}} n_{\rm{d}}, \label{rate_eq}
\end{equation}
where $n_{\rm{a}}$ and $n_{\rm{d}}$ are the density of atoms and that of molecules, respectively, $\alpha$ is the one-body decay rate of a molecule, and $K_{\rm{ad}}$ and $K_{\rm{dd}}$ are the loss coefficients for atom-dimer and dimer-dimer inelastic collisions, respectively. In fitting Eq. (\ref{rate_eq}) to experimental data, we use the peak densities of atoms and molecules for $n_{\rm{a}}$ and $n_{\rm{d}}$. 
We also assume that the spatial distribution of atoms obey an ideal Fermi-Dirac distribution in a harmonic trap, and $n_{\rm{a}}$ is time-independent since the number of atoms far exceeds that of molecules\cite{atom_loss}. For the profile of molecules, we assume a Gaussian spatial distribution $n_{\rm{d}}=N_{\rm{d}}/((2\pi)^{3/2}\sigma_{x} \sigma_{y} \sigma_{z})$, where $\sigma_{x}$, $\sigma_{y}$, and $\sigma_{z}$ are the root-mean square radii of the molecular cloud in the $x$, $y$, and $z$ directions, respectively. Time independent cloud size is assumed in the analysis\cite{Syassen,Mukaiyama}.

\begin{table}[b]
\caption{One-body decay rate $\alpha$, atom-dimer ($K_{\rm{ad}}$) and dimer-dimer ($K_{\rm{dd}}$) inelastic collision coefficients for $|1\rangle - |1\rangle$, $|1\rangle - |2\rangle$, and $|2\rangle - |2\rangle$ molecules.\label{coef_table}}
\vspace{0.5mm}
\begin{tabular}{c|c|c|c}
\hline\hline
 &  $\alpha$ (s$^{-1}$)  &  $K_{\rm{ad}}$($10^{-11}$cm$^3$s$^{-1}$)  &  $K_{\rm{dd}}$($10^{-10}$cm$^3$s$^{-1}$) \\ \hline
$|1\rangle - |1\rangle$  & 0 & $2.4^{+0.5}_{-0.3}$ &  2.8 $\pm$ 0.3\\
$|1\rangle - |2\rangle$  &  260 $\pm$ 150  &  $6.8^{+1.5}_{-1.1}$ & 8.1 $\pm$ 1.1\\
$|2\rangle - |2\rangle$  &  480 $\pm$ 60  & - & - \\
\hline\hline
\end{tabular}
\end{table}

The measured decay rates are summarized in Table \ref{coef_table}.
For $|1\rangle - |1\rangle$ molecules with no unpaired atoms, the decay curve is well fitted by Eq. (\ref{rate_eq}) without the one-body decay term.
As mentioned above, the molecular cloud undergoes breathing-mode oscillations, and the molecular density changes in time up to 20~ms. Therefore, we fit the decay data to extract $K_{\rm{dd}}$ only at $t \ge 20$~ms, until when the molecular cloud is expected to reach thermal equilibrium. 
For the $|1\rangle - |2\rangle$ molecules alone in the trap, we used the whole data for the fit unlike the case of $|1\rangle - |1\rangle$ molecules because of the short lifetime. Since $|2\rangle$ is not the lowest energy state, $|1\rangle - |2\rangle$ molecules are expected to undergo dipolar decay into the lower spin state.
Therefore, both $\alpha$ and $K_{\rm{dd}}$ contribute to the decay. With the initial density $n_{\rm{d}}=4.8 \times 10^{11}$~$\rm{cm}^{-3}$, the decay rate $K_{\rm{dd}} n_{\rm{d}}$ is evaluated to be 390~Hz, which is 15 times larger than the one-body decay rate $\alpha$.
For $|2\rangle - |2\rangle$ molecules alone in the trap, an exponential decay fits the data well, indicating that the one-body decay is dominant.
While we are able to extract $K_{\rm{ad}}$ for the $|1\rangle - |2\rangle$ molecules from the decay data, it is difficult to extract $K_{\rm{ad}}$ for $|2\rangle - |2\rangle$ molecules due to the low signal-to-noise ratio. Here, $n_{\rm{a}}$ is estimated to be 4.5, 8.0, $4.5 \times 10^{12}$~cm$^{-3}$ for Fig. \ref{lifetime}(a)-\ref{lifetime}(c), respectively. $n_{\rm{d}}$ is estimated to be 0.9, 4.8, 1.1~$\times 10^{11}$~cm$^{-3}$ for $|1\rangle - |1\rangle$ (at $t=20$~ms), $|1\rangle - |2\rangle$ (at $t=0$~ms), $|2\rangle - |2\rangle$ (at $t=0$~ms), respectively.

\begin{figure}[t]
\includegraphics[scale = 0.5]{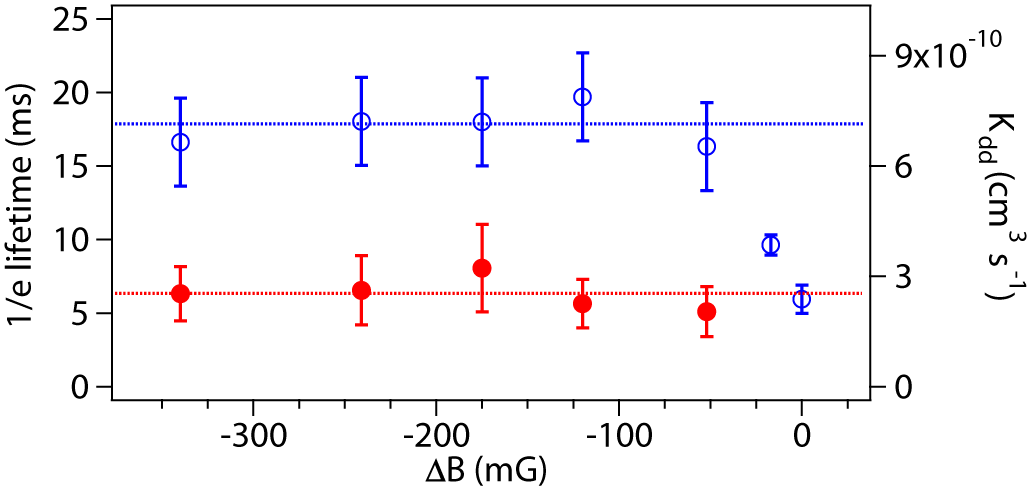}
\caption{Lifetime (open circles) and dimer-dimer inelastic collision coefficient (filled circles) for $|1\rangle - |1\rangle$ molecules vs magnetic-field detuning. The dotted (dashed) lines show the averages of $K_{\rm{dd}}$ ($1/e$ lifetime) for $\Delta B < -50 \rm{mG}$.
\label{inelastic_coef}}
\end{figure}

We measured the $\Delta B$ dependence of the $1/e$ lifetime and $K_{\rm{dd}}$ of the $|1\rangle - |1\rangle$ molecules. The $1/e$ lifetime of molecules is plotted in Fig. \ref{inelastic_coef} (open circles). For $-340 \rm{mG} < \Delta B < -50 \rm{mG}$, the lifetime is independent of $\Delta B$ and similar to the thermalization time of 20~ms. The filled circles in Fig. \ref{inelastic_coef} show $K_{\rm{dd}}$ evaluated from the decay curve in the same manner as in Fig. \ref{lifetime}(a).
Since the lifetime is expected to become shorter when the magnetic field crosses the resonance, our uncertainty in the resonance position could be as large as 50~mG.
We evaluate and plot $K_{\rm{dd}}$ in Fig. \ref{inelastic_coef} only in the range between $-340$~mG and $-50$~mG because neglecting the one-body decay term is valid only in the bound-molecular side of the Feshbach resonance.

In conclusion, we studied the elastic and inelastic collisional properties of $^6$Li$_2$ $p$-wave Feshbach molecules.
From the thermalization time of the breathing-mode oscillation, we evaluated the dimer-dimer elastic collision rate.
From the loss measurements expressed as a function of the hold time, the dimer-dimer inelastic loss coefficient for $|1\rangle - |1\rangle$ and $|1\rangle - |2\rangle$ molecules and the one-body decay coefficients for $|1\rangle - |2\rangle$ and $|2\rangle - |2\rangle$ molecules were obtained. We also evaluated the atom-dimer inelastic collision rates for $|1\rangle - |1\rangle$ and $|1\rangle - |2\rangle$ molecules. In the present experiment, we created $7 \times 10^3$ molecules at a temperature of 1~$\mu$K after thermalization. Provided molecules are distributed equally among three different $m_l$ states, the phase space density of the molecular gas is estimated to be $4 \times 10^{-3}$. 
Our measurements show that the ratio of elastic to inelastic collision rate is about five, indicating that a conventional evaporative cooling would not be efficient\cite{Ketterle}. To overcome this, one could use Feshbach resonance to enhance the dimer-dimer elastic collisions. One can also utilize the optical lattice to single out a specific $m_l$ state\cite{Gunter} which should have a higher ratio of elastic to inelastic collision rate.



\begin{thebibliography}{99}
\bibitem{atom_loss} The effect of atomic loss is estimated and taken into account as errors of $K_{\rm{ad}}$ in Table \ref{coef_table}.
\end{thebibliography}
\end{document}